# Features

## An Area Efficient 2D Fourier Transform Architecture for FPGA Implementation

Atin Mukherjee[1] and Debesh Choudhury[2]
[1]Department of Electronics and Communication Engineering, Dr. B.C Roy Polytechnic, Durgapur 713206, India
[2]Infosensys Consulting, Dakshinayan, PO – Sodepur, Kolkata 700110, West Bengal, India

**Abstract** – Two-dimensional Fourier transform plays a significant role in a variety of image processing problems, such as medical image processing, digital holography, correlation pattern recognition, hybrid digital optical processing, optical computing etc. 2D spatial Fourier transformation involves large number of image samples and hence it requires huge hardware resources of field programmable gate arrays (FPGA). In this paper, we present an area efficient architecture of 2D FFT processor that reuses the butterfly units multiple times. This is achieved by using a control unit that sends back the previous computed data of N/2 butterfly units to itself for $(\log_2 N - 1)$ times. A RAM controller is used to synchronize the flow of data samples between the functional blocks. The 2D FFT processor is simulated by VHDL and the results are verified on a Virtex-6 FPGA. The proposed method outperforms the conventional $N \times N$ point 2D FFT in terms of area which is reduced by a factor of $\log_N 2$ with negligible increase in computation time.

### Introduction

Fourier Transform is one of the most widely used operations in Digital Signal Processing and plays a significant role in many signal processing applications such as image processing, diffraction, propagation, holography, fiber optics, lasers etc. Basically spatial Fourier transform is used to get the information about phase and magnitude of any spatial signal such as image [1]. A spatial signal contains lots of information about itself. Therefore, to calculate the sine and cosine component of a spatial signal requires lots of hardware resources. Simultaneously, the area and power consumption are directly proportional to number of hardware resources. Though large number of computational blocks may increase the speed of operation but also increase the area and power dissipation of the processor [2,3]. In today's world, large area and high power dissipation are two major drawbacks of a system.

Fully spatial parallel FFT architecture also known as array architecture [4], based on the complete Cooley-Tukey algorithm layout, is one of the potential high throughput designs. However, the implementation of the array architecture is hardware intensive. It achieves high performance by using spatial parallelism, while requiring more routing resources. However, as the problem size grows, unfolding the architecture spatially is not feasible due to serious power and area issue brought by complex interconnections.

The column architecture [5,6,7] uses an approach that requires less area on the chip than the array architecture. It is done by collapsing all the columns in array architecture into one column; hence a new frame cannot be processed until the processing of the current frame is finished. Hence, this architecture is not suitable for pipelining. The area requirement is obviously smaller, only $N/r$ radix-r elements, than for the array architecture.

The Pipelined architectures [8,9,10] are useful for FFTs that require high data throughput. The basic principle with pipelined architectures is to collapse the rows, instead of the stages like in column architectures. The architecture is built up from radix butterfly elements with commutators in between. Radix-2 Multi-path Delay Commutator [11,12] was probably the most popular approach for pipeline implementation of radix-2 FFT algorithm.

In this paper, we propose an area efficient architecture of 2D Fourier transform by reusing $N/2$ numbers of butterfly units instead of $\frac{N}{2}\log_2 N$ butterfly units for each 1D FFT architecture. This is achieved by a control unit (CU) which sends back the previous computed data of $N/2$ butterfly units to itself for $(\log_2 N - 1)$ times [13]. A RAM Control unit is used to synchronize the dataflow for both 1D FFT block and controls the internal data samples to prevent data loss for fast processing. The area requirement is obviously smaller, only $N/2$ radix-2 elements, than the array architecture and pipelined architectures for each 1D FFT architecture, N being the number of sample points.

## 1. Traditional 2D FFT Algorithm

The Cooley-Tukey one dimensional (1D) FFT algorithm is the most commonly used for calculating the frequency components of one dimensional sequence $x(n)$. This algorithm uses a recursive way of solving FFT of any arbitrary size $N$ of input sequence. The technique divides the larger FFT into smaller FFT, which subsequently reduces the complexity of the algorithm. If the size of the FFT is $N$ then this algorithm makes $N = N1 \cdot N2$ where $N1$ and $N2$ are smaller FFT's. Radix-2 decimation in time (DIT) divides the size N FFT's into two interleaved FFT's of size $\frac{N}{2}$. A FFT of N-point discrete-time complex sequence $x(n)$, indexed by $n = 0,1, \ldots (N-1)$ is defined as:

$$Y(k) = \sum_{n=0}^{N-1} x(n) W_N^{nk}, k = 0,1, \ldots N-1 \quad \text{Where } W_N = e^{-j2\pi/N} \quad (1)$$

From the equation (1) we can drive the Radix-2 2D Fast Fourier Transform algorithm. The 2D FFT of a space limited 2D function $F(x,y)$ can be expressed as

$$F(u,v) = \sum_{x=0}^{N-1} \sum_{y=0}^{M-1} F(x,y) W_N^{ux} W_M^{vy} \quad (2)$$

Where $F(x,y)$, is an input image and $F(u,v)$ is an output image. Size of input image $N \times M$. From the equation (2) it is clear that we are performing 1D FFT operation for two times to generate the 2D FFT output. For a square image $N = M$, length of 1D FFT operation is same for both 1D FFT blocks. Figure 1 shows the block diagram of 2D fourier transform.

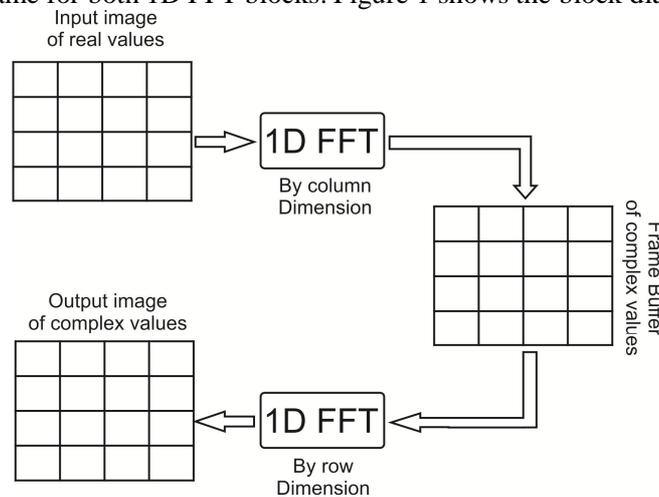

Figure 1: Block diagram of 2D Fourier Transform

## 2. Proposed 2D FFT Algorithm

From the figure 1, it is clear that to perform 2D FFT operation we have to compute one dimensional FFT operation for two times. In our proposed 2D FFT architecture, we reduced the area of these two 1D FFT block drastically. These 1D FFT blocks uses only $\frac{N}{2}$ number of butterfly units (BU) and reused for $(\log_2 N - 1)$ times [13]. Therefore, area of each 1D FFT blocks drastically reduced by a factor of $\log_N 2$. Therefore, the proposed architecture of 2D FFT processor requires $BU_{Proposed2DFFT}$ number units which is given by

$$BU_{Proposed2DFFT} = 2 \times \frac{N}{2} = N \quad (3)$$

Therefore, for a traditional 2D FFT processor, the total number of butterfly units is given by

$$BU_{Traditional2DFFT} = 2\left(\frac{N}{2}\right) \log_2 N = N \log_2 N \quad (4)$$

The proposed architecture of 2D FFT processor reduces the number of butterfly units by a factor of $\alpha_{2D}$, which given by

$$\alpha_{2D} = \frac{N}{N \log_2 N} = \log_2 N^{-1} = \log_N 2 \quad (5)$$

Table 1. shows that the number of butterfly unit (BU), multiplier and adder/subtractor for the proposed 2D FFT is less compared to that of the traditional 2D FFT. Figure 2 shows a comparative of hardware resources between proposed 2D FFT architecture and traditional 2D FFT architecture.

TABLE 1.
COMPARISON OF BUTTERFLY UNITS, MULTIPLIER AND ADDER/SUBTRACTORS FOR 2D FFT PROCESSOR

|  | Traditional 2D FFT | Proposed 2D FFT |
|---|---|---|
| Butterfly Unit (BU) | $N \log_2 N$ | $N$ |
| Multiplier | $N \log_2 N$ | $N$ |
| Adder/subtractor | $2N \log_2 N$ | $2N$ |

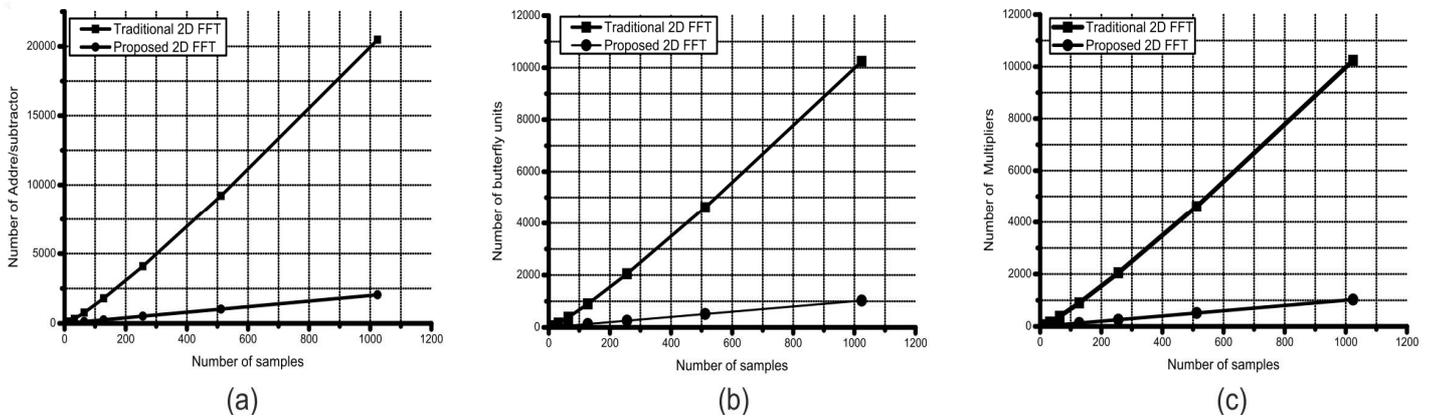

Figure 2: Comparison of number of (a) adder/subtractor, (b) Butterfly units and (c) multiplier required in between traditional and proposed 2D FFT processor.

## 3. Proposed Hardware Architecture of 2D FFT Processor

Figure 3 shows the proposed hardware architecture of the 2D FFT processor. In this architecture two 1D FFT blocks [13] are working simultaneously. The first 1D FFT block takes input samples from the external world and after FFT computation the results are stored either into the RAM1 or RAM2. Now the second 1D FFT block takes input either from RAM1 or RAM2 and generates the final 2D FFT output samples. These RAM units are basically utilized to store the output samples of the first 1D FFT blocks and the RAM units are of same size. The selection of the RAM units are done by the RAM controller.

At the initial state, the RAM controller is reset to 'sel' ($sel = 0$ $and$ $\overline{sel} = 1$) output line. Then, the output samples of the first 1D FFT block will be stored into the RAM1 and the second 1D FFT block takes input samples from RAM2. But when both the RAM units reaches its maximum limit, the RAM controller simply inverts the 'sel' ($sel = 1$ $and$ $\overline{sel} = 0$) output line. Then, the output samples of the first 1D FFT block are stored into RAM2 and the second 1D FFT block takes input samples from RAM1. Again, when both the RAM units reaches its maximum limit, the RAM controller again inverts the 'sel' ($sel = 0$ $and$ $\overline{sel} = 1$) output line. In this way, the complete 2D FFT computation will be performed continuously. The timing diagram of 'sel' line is shown in figure 4.

When one RAM unit is busy in write operation, at that time another RAM unit is busy in read operation and the RAM controller plays a very vital role to synchronise the read and write operation of the RAM units. RAM controller generates all the output signal depends on the DONE signal of 1D FFT block. DONE signal goes to HIGH state for one clock period when output samples of 1D FFT blocks are valid. Otherwise DONE signal always maintain LOW state. The behaviour of DONE signal is same with the Output Select Line (OSL) of 1D FFT block. The detailed architecture and behaviours of 1D FFT blocks are explained in the next subsections.

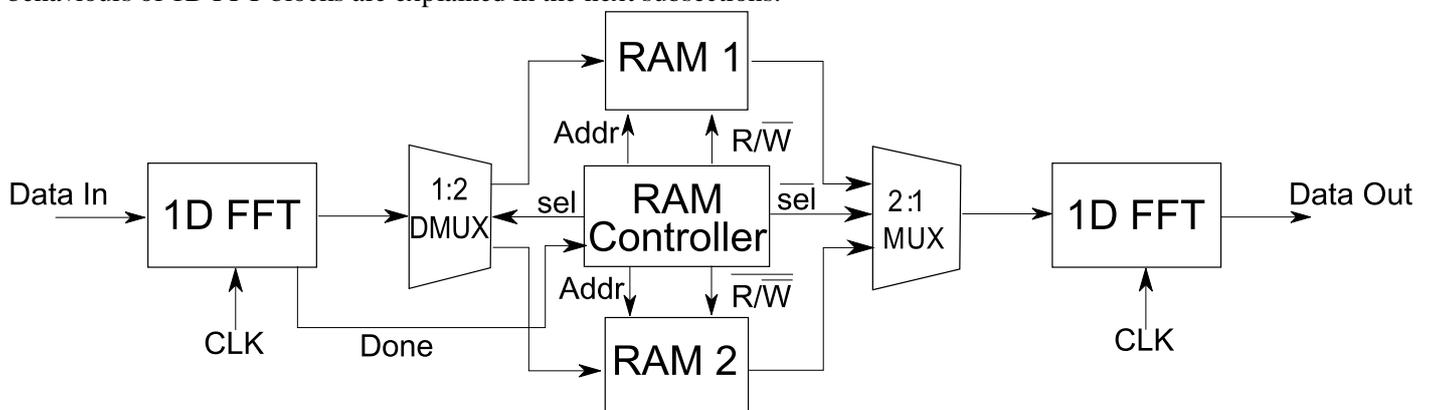

Figure 3: Architecture of proposed 2D FFT processor

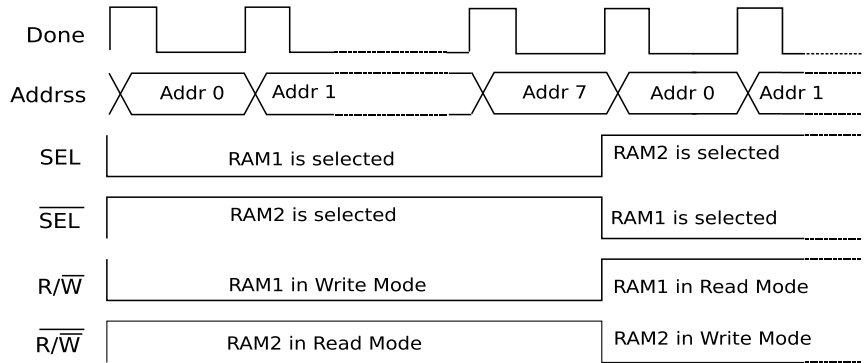

Figure 4: Timing diagram of RAM Controller Block

A. *Architecture of 1D FFT processor*

The major advantage of this 1D FFT architecture is that it is a very area efficient architecture. 1D FFT processor is designed based on reutilization of butterfly units [13]. Area of 1D FFT processing unit is reduced by reusing the $N/2$ number of butterfly units for $(\log_2 N - 1)$ times, where $N$ is the number of input samples. Therefore, each 1D FFT processor requires $N/2$ number of butterfly units instead of $\left(\frac{N}{2} log_2 N\right)$ numbers. Table 2 shows a comparative study of hardware resources between traditional FFT architecture and proposed 1D FFT architecture. Figure 5 shows the complete architecture of 1D FFT processor. This architecture consists of Butterfly units, control unit and routing network.

TABLE 2.
COMPARISON OF BUTTERFLY UNITS, MULTIPLIER AND ADDER/SUBTRACTORS FOR 1D FFT PROCESSOR

|  | Traditional 1D FFT | Proposed 1D FFT |
|---|---|---|
| **Butterfly Unit (BU)** | $\frac{N}{2} log_2 N$ | $\frac{N}{2}$ |
| **Multiplier** | $\frac{N}{2} log_2 N$ | $\frac{N}{2}$ |
| **Adder/subtractor** | $N log_2 N$ | $N$ |

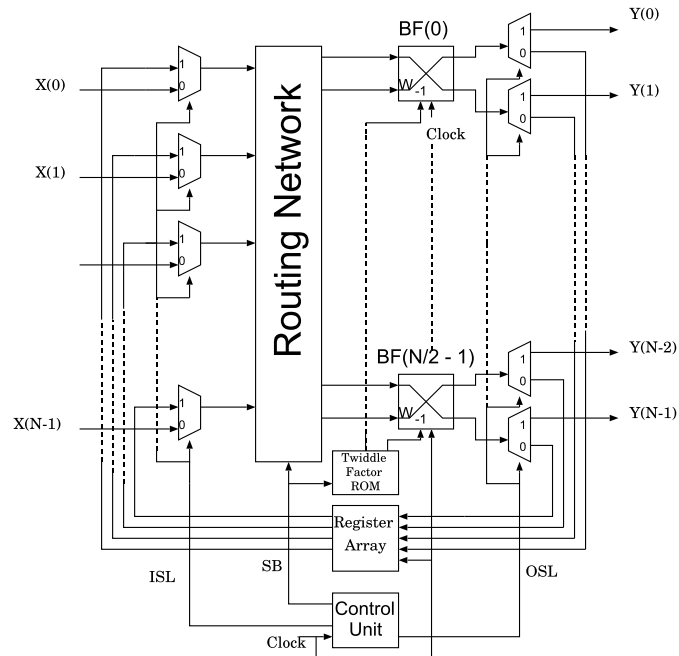

Figure 5: Architecture of 1D FFT processor

Butterfly units are the main computational block for the FFT operations. In figure 6a shows the mathematical model of a butterfly unit. In figure 6a, A and B are the odd and even input samples respectively. Even samples are multiplied by twiddle factor co-efficient ($W_N$). The result of multiplication is subtracted from the odd sample to generate the even output samples and added with the same odd sample to generate the odd output samples. To generate these odd and even output samples the butterfly unit (BU) needs only one clock cycle. The complete timing diagram of the butterfly unit is shown in the figure 6b.

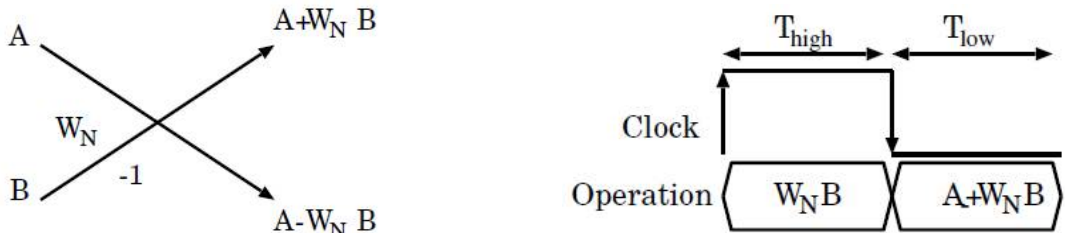

Figure 6: a) Mathematical model of butterfly units. b) Timing diagram of butterfly units

Routing Network module shuffles the input samples according the stage of computation and supplies the appropriate samples to the butterfly units. The control unit counts the number of stages and supplies the number of current stage through Stage Bus (SB). Depending on the value of SB, the routing network module shuffles the input samples. The register array is basically used to store the previous output samples of the butterfly units.

Here the control unit plays a vital role to synchronize all the blocks of the 1D FFT processors. Depending on the input clock signal, it counts the number of stages and controls the input, output and feedback data path. Table 3 shows all input and output signal descriptions of the control unit.

TABLE 3.
INPUT AND OUTPUT SIGNAL DESCRIPTION OF CONTROL UNIT

| Signal name | Signal type | Description |
|---|---|---|
| Clock (CLK) | Input | System clock input |
| ISL | Output | **Input select line**, 1 bit output line. If ISL is in reset condition then 1D FFT processor takes input from the external world. Otherwise, for set condition processor takes previous output from the register array through internal feedback data path. |
| OSL | Output | **Output select line**, 1 bit output line. If OSL is in set condition then output of butterfly unit will go to the external world as a final FFT output. Otherwise, for reset condition the output of butterfly units will store into the register array. |
| SB | Output | **Stage Bus** generates the number of current stage. Control unit increment the stage by 1 for every rising edge of clock. Control unit starts count from 0 to $\log_2 N - 1$. Stage bus width depends on the number of stages. |

At initial stage, **Input Select Line** (ISL) goes to LOW state ($ISL = '0'$) to select the data from the external world. After completing the computation of first stage the output of the butterfly unit should forward for next stage. Therefore, the **Output Select Line** (OSL) goes to LOW state ($OSL =' 0'$) and the output samples of first stage are stored into the register array. Now for the next stage the Input Select Line (ISL) goes to HIGH state ($ISL =' 1'$) to select the data from the register array. After the first stage the ISL will maintain the HIGH stage for remaining ($\log_2 N - 1$) stages and OSL will go to the HIGH state at the time of computation of last stage ($\log_2 N$th). Figure 6 shows the timing diagram of 8-point 1D FFT processor where $X(n)$ and $Y(K)$ denotes input and output samples and $f(K)$ are the output samples of the previous stage, which is stored into the register array. Figure 7 shows the complete timing diagram of 8-point FFT processor and figure 8 shows the RTL diagram of 8-point 1D FFT processor. Table 4 shows FPGA resource utilized summary of proposed 1D FFT processor.

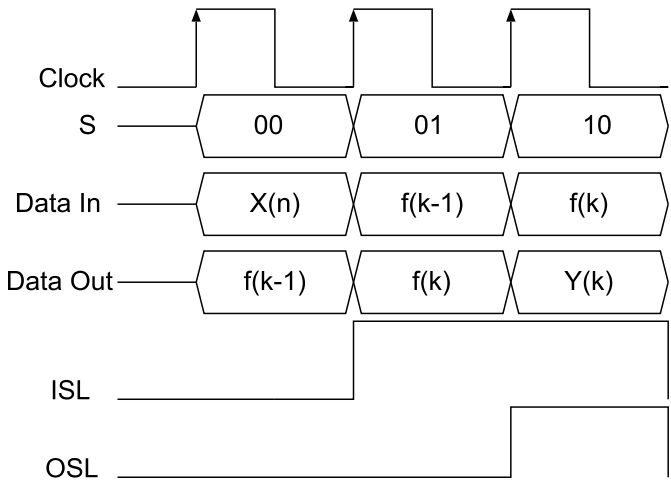

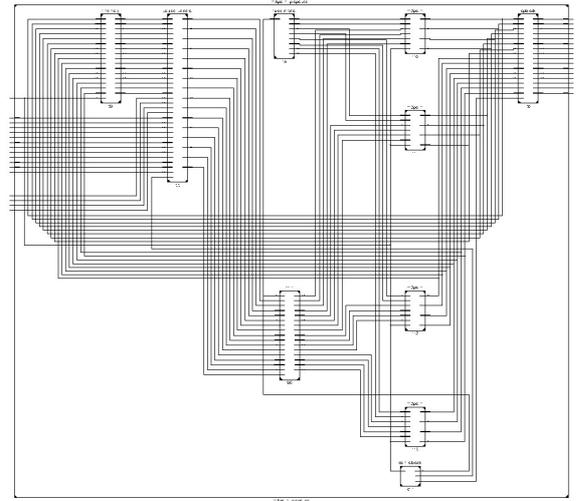

Figure 7: Timing diagram of 8-point FFT processor

Figure 8. RTL diagram of 8-point 1D FFT processor

TABLE 4:
DEVICE UTILIZATION SUMMARY OF PROPOSED 1D FFT PROCESSOR

| Selected Device | 6vsx475tff1759-2 |
|---|---|
| Number of Slice Registers | 301 |
| Number of Slice LUTs | 748 |
| Number of DSP48E1s | 16 |

## 5. Implementation and Results

The proposed architecture of 2D FFT processor is coded using VHDL, emulated and synthesized using Xilinx ISE 14.2. Table 5 presents a comparative study of FPGA resource utilization summary for 8x8 2D FFT processor. The device resource utilization summary shows that the proposed 2D FFT architecture requires very less number of resources such as Slice, LUT, DSP blocks etc. The utilized FPGA resource count has been defined as the design area [14,15]. Number of utilized recourses are directly proportional to the design area. Table 6 shows the comparison of timing delay between the traditional and proposed 8x8 2D FFT processor where processing speed of proposed architecture is negligibly higher than traditional architecture. Estimated power consumption of the proposed 2D FFT processor has been measured using Xilinx XPower Analyzer tool and turns out as 6.910W.

TABLE 5:
DEVICE UTILIZATION SUMMARY BETWEEN PROPOSED 2D FFT PROCESSOR AND TRADITIONAL 2D FFT PROCESSOR

| Device utilization summary | | |
|---|---|---|
| | Proposed 2D FFT | Traditional 2D FFT |
| Selected Device | 6vsx475tff1759-2 | 6vsx475tff1759-2 |
| Number of Slice Registers | 1132 | 2603 |
| Number of Slice LUTs | 2109 | 5327 |
| Number of DSP48E1s | 33 | 97 |

TABLE 6:
COMPARISON OF DELAY BETWEEN TRADITIONAL AND PROPOSED 2D FFT PROCESSOR

| Algorithm | Delay(nsec) |
|---|---|
| Traditional 2D FFT | 32.129 |
| Proposed 2D FFT | 32.487 |

## 5. Conclusion

In conclusion, we have presented a novel architecture of area efficient Radix-2 2D FFT processor. The proposed architecture is area efficient over the traditional 2D FFT architecture. Reutilization of two-point butterfly units of single stage helps reducing the hardware area significantly. The proposed 2D FFT architecture has been simulated using Xilinx ISE 14.2 and synthesized for Xilinx Virtex-6 FPGA. The performance analysis has been carried out in terms of overall response time and utilization of hardware resources of FPGA. The detailed analysis reveals that the proposed architecture offers drastic hardware area reduction compared to traditional 2D FFT processors without increase in overall response time. Further improvements can be achieved by designing silicon layout and analyzing the post-layout performance trade-offs.

**About the Authors**

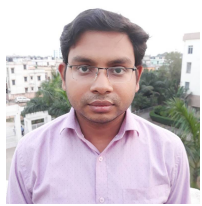
**Atin Mukherjee** (mukherjee.atin1@gmail.com) is a Lecturer in the Department of Electronics and Telecommunication Engineering at Dr. B.C. Roy Polytechnic, Durgapur. He has 6 years of research experience in the areas of DSP, VLSI, and Embedded Systems. His current research projects include hardware and software design for video image processing using open source software.

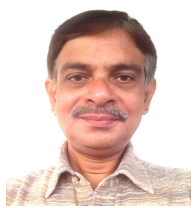
**Debesh Choudhury** (debesh@iitbombay.org) is a technology developer with 28+ years of research and 15+ years of teaching experience. He earned a PhD from IIT Bombay, and had two years postdoctoral research stint at the University of Electro-Communications, Tokyo. His current research projects include Biometric Security, Privacy Protection, 3D Sensing, Artificial Color Perception, and Internet of Everything (IoE). He is an active user and trainer of GNU/Linux and FOSS. He is also a musician and a social media marketer. He was featured twice in Forbes Entrepreneurs channel. He is a Senior Member of IEEE and SPIE. More details in https://www.linkedin.com/in/debeshchoudhury/